\documentclass[11pt]{article}
\usepackage[a4paper]{geometry}

\usepackage{amsmath, amsthm, amssymb}
\usepackage{graphicx}
\usepackage[round]{natbib}
\usepackage{url}
\usepackage[hidelinks,colorlinks=true,linkcolor=blue,citecolor=blue,urlcolor=blue]{hyperref} 
\usepackage{booktabs}
\usepackage{subcaption}
\usepackage{placeins} % Simple \FloatBarrier for reasonable plot placement

\usepackage{fancyvrb}
\DefineVerbatimEnvironment{CodeInput}{Verbatim}{fontshape=sl}
\DefineVerbatimEnvironment{CodeOutput}{Verbatim}{}
\newenvironment{CodeChunk}{}{}

\newcommand{\balnet}{\texttt{balnet}}
\newcommand{\cvbalnet}{\texttt{cv.balnet}}
\newcommand{\adelie}{\texttt{adelie}}
\newcommand{\glmnet}{\texttt{glmnet}}

\newcommand{\CBPS}{\texttt{CBPS}}
\newcommand{\RCAL}{\texttt{RCAL}}
\newcommand{\ebal}{\texttt{ebal}}
\newcommand{\sbw}{\texttt{sbw}}
\newcommand{\survey}{\texttt{survey}}

\newcommand{\Eigen}{\texttt{Eigen}}
\newcommand{\Rcpp}{\texttt{Rcpp}}
\newcommand{\Rlang}{\texttt{R}}

\newcommand{\cpp}{\texttt{C++}}

\newcommand{\code}[1]{\texttt{#1}}

\newcommand{\EE}[2][]{\mathbb{E}_{#1}\left[#2\right]}
\newcommand{\PP}[2][]{\mathbb{P}_{#1}\left[#2\right]}
\newtheorem{prop}{Proposition}

\author{
  Erik Sverdrup\textsuperscript{1} \hspace{2em}
  Trevor Hastie\textsuperscript{2}  \\[0.2cm]
  \small \textsuperscript{1}Department of Econometrics \& Business Statistics, Monash University\\
  \small \textsuperscript{2}Department of Statistics, Stanford University
}

\date{}
\title{\balnet:~Pathwise Estimation of Covariate\\ Balancing Propensity Scores}

\begin{document}
\maketitle

\begin{abstract}
We present \balnet, an \Rlang~package for scalable pathwise estimation of covariate balancing propensity scores via logistic covariate balancing loss functions.
Regularization paths are computed with \citet{yang2024fast}'s generic elastic net solver, supporting convex losses with non-smooth penalties, as well as group penalties and feature-specific penalty factors.
For lasso penalization, \balnet~computes a regularization path of balancing weights from
the largest observed covariate imbalance to a user-specified fraction of this maximum.
We illustrate the method with an application to spatial pixel-level balancing for constructing synthetic control weights for the average treatment effect on the treated, using satellite data on wildfires.
\end{abstract}

\section{Introduction} \label{sec:intro}

The propensity score plays a central role in estimating causal effects in observational settings \citep{rosenbaum1983central}.
It is defined as the probability that an individual with a given pre-treatment covariate profile is assigned to the treatment, and is frequently estimated using logistic regression. 
In high-dimensional settings, such as applications with many confounders, rich basis expansions, or higher-order interactions, a standard approach is to estimate the propensity score using lasso \citep{tibshirani1996regression} or elastic net penalized logistic regression \citep{zou2005regularization}, as implemented in the widely used \Rlang~package \glmnet~\citep{glmnet}.

A primary use of the propensity score is inverse probability weighting (IPW), where outcomes are reweighted to construct treated and control groups that are comparable in terms of their pre-treatment covariates.
Under the true (oracle) propensity score, IPW exactly balances covariate means between treatment arms and the overall population.
However, even without model misspecification, this exact balancing property does not generally hold for propensity scores estimated using standard maximum likelihood logistic regression, which prioritizes accurate prediction of treatment assignment, rather than covariate balance.
\citet{imai2014covariate} introduced the notion of covariate balancing propensity scores, propensity score estimators explicitly designed to optimize covariate balance. Causal effects estimated using these balancing weights can be more robust to model misspecification and more efficient than approaches that estimate propensity scores via maximum likelihood \citep{ben2021balancing, chattopadhyay2020balancing}.
These weights can also improve doubly robust methods, such as AIPW \citep{robins1994estimation}, where the IPW weights act as debiasing weights \citep{van2024doubly, vermeulen2015bias}.

A number of methods have been proposed to directly target covariate balance.
These include the covariate balancing propensity score estimator, \CBPS~\citep{CBPS}, which enforces moment conditions via generalized method of moments,
and entropy balancing, \ebal~\citep{hainmueller2012entropy}, which directly optimizes weights to achieve balance. 
Such methods may achieve exact balance in the relatively small datasets commonly encountered in the social sciences. However, in large-scale datasets, these approaches can be computationally challenging to scale and are often infeasible, as exact balance may not be attainable\footnote{A closely related approach from the survey literature, available in the \survey~package \citep{lumley2004analysis}, is raking, which calibrates survey weights so that sample marginal distributions match known population margins.}.
The stable balancing weights approach, \sbw~\citep{zubizarreta2015stable}, relaxes the requirement of exact balance and instead optimizes weights to satisfy balance levels pre-specified by the researcher. 
While this approach is effective, it requires the user to choose these levels a priori. If the chosen level is infeasible for a given dataset, or if a different bias–variance tradeoff is desired, the underlying optimization problem must be re-solved. In practice, this leads to iterative tuning between balance constraints and downstream model validation \citep{keele2025balancing}.

Motivated by these challenges, we seek an algorithmic solution that simplifies the selection of balance levels in a manner analogous to regularization paths for penalized generalized linear models. Our approach builds on recent work that unifies many balance-oriented estimators through covariate balancing loss functions \citep{zhao2019covariate}.

For the logistic link, \citet[Chapter 7]{wagerbook} note (emphasis added):
\begin{quote}
    $\dots$if we believe in a linear-logistic specification and want to use an IPW estimator, then we should learn the propensity model by minimizing the \emph{covariate-balancing loss} function rather than by the usual \emph{maximum likelihood loss} used for logistic regression.
\end{quote}
This observation motivates the use of covariate balancing loss functions with modern regularization techniques. The penalized versions of these loss functions correspond to the primal formulation of optimization problems that constrain covariate imbalance \citep{ben2021balancing, tan2020regularized, zhao2019covariate}. As a consequence, the interpretation of the penalty parameter differs fundamentally from that in penalized logistic regression, where it acts as a coefficient budget. For lasso in particular, the penalty parameter directly controls the maximum allowable absolute imbalance across covariates, offering an intuitive metric to diagnose overlap issues, as well as monitor solver progress in applications.

To our knowledge, the closest existing software is \RCAL~\citep{RCAL}, which implements lasso regularization for covariate balancing loss functions. However, because this solver is written entirely in \Rlang, it can be difficult to scale to large datasets. This limitation is particularly salient in large-scale applications such as spatial or pixel-level balancing, where both the number of observations and covariates are large. 

The package \balnet~for \Rlang~\citep{Rcore}, available at \href{https://CRAN.R-project.org/package=balnet}{CRAN.R-project.org/package=balnet}, is designed to address these limitations by combining covariate balancing loss functions with modern optimization techniques for large-scale regularization problems. Our approach leverages proximal quasi-Newton methods for generic convex losses that incorporates standard algorithmic refinements for coordinate descent such as screening and active-set rules \citep{glmnet, simon2013blockwise, tibshirani2012strong}. The core implementation uses the \adelie~\cpp~elastic net solver by \citet{yang2024fast}, interfaced with \Rlang~via \Rcpp~\citep{eddelbuettel2011rcpp}, and relies on numerical linear algebra routines from \Eigen~\citep{eigenweb}. In an application from \citet{wu2023low} involving spatial balancing with about 140,000 observations and 500 covariates, \balnet~computes the full regularization path, from raw, unweighted imbalance down to commonly accepted balance thresholds, within minutes on a standard laptop.

\section{Regularization paths for covariate balancing propensity scores} \label{sec:cbps}

Let $Y_i(1)$ and $Y_i(0)$ denote the potential outcomes under treatment and control, respectively \citep{imbens2015causal}. The realized treatment assignment is $W_i \in \{0,1\}$ and the observed outcome is $Y_i = Y_i(W_i)$, for a sample of units $i = 1, \ldots, n$. In observational settings, a common identifying assumption is unconfoundedness: conditional on a set of pre-treatment covariates $X_i \in \mathbb{R}^p$, treatment assignment is as good as random. Under this assumption, a central object is the propensity score,
\[
e(x) = \PP{W_i = 1 \mid X_i = x}.
\]

Given suitable overlap conditions, potential outcome means are identified via inverse-propensity weighted outcomes. We begin by considering the treated mean, extensions to the average treatment effect (ATE) and the average treatment effect on the treated (ATT) follow by symmetry. The treated mean is identified by
\[
\EE{Y_i(1)} = \EE{\frac{W_i Y_i}{e(X_i)}}.
\]
The oracle propensity score also satisfies population-level covariate balance,
\[
\EE{\frac{W_i X_{ij}}{e(X_i)}} = \EE{X_{ij}}, \quad j = 1, \ldots, p.
\]
For notational simplicity, we describe the setting in terms of linear dependence on $X_i$, the discussion extends directly to nonlinear specifications via basis expansions and interactions.

Under a logistic propensity score model,
\[
e(x) = \frac{1}{1 + \exp\left(-(\beta_0 + x^\top \beta)\right)}.
\]
If $\beta$ is estimated by maximum likelihood using the standard Bernoulli log-likelihood, the gradient equations are
\begin{equation} \label{eq:ML_score}
    \frac{1}{n} \sum_{i=1}^{n} \Big(W_i - \hat e(X_i)\Big) X_{ij} = 0,
    \quad j = 1, \ldots, p,
\end{equation}
where $\hat e(X_i)$ are the fitted propensity scores. Maximum likelihood essentially balances predicted and observed treatment. Achieving finite-sample covariate balance using the fitted propensities requires that
\begin{equation} \label{eq:balance_mu1}
    \frac{1}{n} \sum_{i=1}^{n} \frac{W_i X_{ij}}{\hat e(X_i)} = \frac{1}{n} \sum_{i=1}^{n} X_{ij}, 
    \quad j = 1, \ldots, p.
\end{equation}
To enforce this covariate balancing condition, we instead work backwards and require the gradients satisfy
\begin{equation}
    \frac{1}{n} \sum_{i=1}^{n} \left( \frac{W_i}{\hat e(X_i)} - 1 \right) X_{ij} = 0, 
    \quad j = 1, \ldots, p.
\end{equation}
Substituting in for the logistic propensity scores, this is equivalent to requiring that $\hat\beta$ satisfies
\begin{equation} \label{eq:balance_logistic}
    \frac{1}{n} \sum_{i=1}^{n} \Bigg( W_i\left[1 + \exp\bigl(-(\hat\beta_0 + X_i \hat\beta)\bigr)\right] - 1 \Bigg)X_{ij}
      = 0,
    \quad j = 1, \ldots, p.
\end{equation}
The loss function that enforces \eqref{eq:balance_logistic} is
\begin{equation} \label{eq:cbps_mu1}
    l_1(\eta) = \frac{1}{n} \sum_{i=1}^{n} \Big( W_i \exp(-\eta_i) + (1 - W_i)\eta_i \Big),
\end{equation}
where $\eta_i = \beta_0 + X_i \beta$. This follows by setting the gradient of $l_1(\eta)$ to zero, which yields the balance conditions in \eqref{eq:balance_logistic}. 
The IPW estimator constructed using such weights has favorable statistical properties: when the logistic model is correctly specified it is $\sqrt{n}$-consistent with efficient variance \citep[Theorem 7.1]{wagerbook}. 
An equivalent perspective is that we target the Riesz representer for $\EE{Y_i(1)}$, the ratio $W_i/e(X_i)$, rather than estimating $e(X_i)$ via maximum likelihood and inverting the fitted probabilities (e.g., \citet{hirshberg2021augmented}; see also \citet{chernozhukov2022riesz}).

The balance condition for the control group mean is
\begin{equation}
    \frac{1}{n} \sum_{i=1}^{n} \frac{(1 - W_i) X_{ij}}{1 - \hat e(X_i)} = \frac{1}{n} \sum_{i=1}^{n} X_{ij}, 
    \quad j = 1, \ldots, p,
\end{equation}
which yields an analogous loss $l_0(\eta)$ for estimating propensity scores that target $\EE{Y_i(0)}$,
\begin{equation}
    l_0(\eta) = \frac{1}{n} \sum_{i=1}^{n} \Big( (1 - W_i) \exp(\eta_i) - W_i\eta_i \Big).
\end{equation}
For estimating the ATE, $\EE{Y_i(1)} - \EE{Y_i(0)}$, there are $2p$ means to balance.
Since each logistic model has $p$ free parameters, \balnet~thus fits two propensity score models: one targeting balance for the treated mean and one for the control mean, unlike maximum likelihood which fits one set of coefficients\footnote{It is sufficient to implement only the loss in \eqref{eq:cbps_mu1}. By symmetry, propensities targeting $\EE{Y_i(0)}$ are obtained by solving \eqref{eq:cbps_mu1} with inverted treatment indicators $W_i' = 1 - W_i$. 
The propensities targeting $\EE{Y_i(0)}$ can also be used to target the ATT, $\EE{Y_i(1) - Y_i(0) \mid W_i = 1}$, since the losses only differ by a scaling (the former weights the control to match $\frac{1}{n} \sum_{i=1}^{n} X_{i}$ while the latter weights the controls to match $\frac{1}{\sum_{i=1}^{n} W_i} \sum_{i=1}^{n} W_iX_{i}$).}.

In practice, directly minimizing \eqref{eq:cbps_mu1} may not be feasible. Exact balance may be unattainable in settings with high-dimensional covariates ($p > n$) or under limited overlap, potentially producing extreme weights.
The covariate balancing loss imposes stronger requirements than maximum likelihood estimation, which only requires overlap, e.g., the absence of complete separation.

\begin{prop}[Proposition S1, \citet{tan2020regularized}]
The loss $l_1(\eta)$ is strictly convex, bounded below, and has a unique minimizer if and only if the following set is empty:
\[
\left\{
\beta \neq 0 :
\eta_i \geq 0 \text{ if } W_i = 1 \text{ for } i = 1, \ldots, n,
\ \text{and} \
\frac{1}{n} \sum_{i=1}^{n} (1 - W_i)\eta_i \leq 0
\right\}.
\]
\end{prop}

This condition is stricter than logistic separation: it not only rules out a linear hyperplane from perfectly separating treated and controls, but also prevents the weighted control mean from exceeding the treated mean along any linear combination of covariates.
Regularization is therefore key to making this approach practical. We consider the penalized objective
\begin{equation} \label{eq:cbps_penalized}
    l_1(\eta) + \lambda P(\beta),
\end{equation}
where $P(\cdot)$ is a suitable penalty. The default choice in \balnet~is the lasso, $P(\beta) = \sum_{j=1}^{p} |\beta_j|$. This penalty has a particularly convenient interpretation for balancing losses \citep{tan2020regularized, zhao2019covariate}. Exact balance requires
\begin{equation}
\frac{1}{n} \sum_{i=1}^{n} \frac{W_i X_{ij}}{\hat e(X_i)} - \frac{1}{n} \sum_{i=1}^{n} X_{ij} = 0,
\quad j = 1, \ldots, p.
\end{equation}
Under lasso penalization, this constraint is relaxed to a box constraint since the corresponding KKT conditions give
\begin{equation} \label{eq:cbps_kkt}
\left|
\frac{1}{n} \sum_{i=1}^{n} \frac{W_i X_{ij}}{\hat e(X_i)} - \frac{1}{n} \sum_{i=1}^{n} X_{ij}
\right|
\leq \lambda,
\quad j = 1, \ldots, p.
\end{equation}
That is, the regularization path indexed by $\lambda$ yields a sequence of solutions with gradually decreasing maximum imbalance tolerances, corresponding to an $\ell_\infty$ bound on the covariate imbalance vector.

\balnet~minimizes \eqref{eq:cbps_penalized} using the group elastic net solver of \citet{yang2024fast}, which supports lasso, ridge, elastic net, and group penalties, by specifying an extension GLM family in \cpp~for the loss \eqref{eq:cbps_mu1}\footnote{$l_1(\eta)$ is not a canonical GLM loss due to the multiple $W_i$ terms, this is not an issue as the solver only requires the loss to be convex in the linear predictors $\eta$ \citep[Section 4]{yang2024fast}.
However, \adelie~defaults to deviance-based step size scaling, which we disable as the deviance difference between the null and full models can be negative for this loss.
We also modify the GLM solver to track balance metrics in place of deviance for interactive path progression.}.
Group penalties, optionally combined with penalty factors, allow \balnet~to balance sets of related covariates jointly, such as interactions or categorical expansions, an approach that is useful in many applied settings \citep[e.g.,][]{ben2021balancing}.

Pathwise optimization is a natural fit for this loss and can provide a useful practical diagnostic for overlap. \balnet~constructs a log-spaced $\lambda$ sequence in the spirit of \glmnet, starting at the smallest $\lambda^{\max}$ for which the solution has $\hat \beta = 0$, and proceeding to a fixed fraction of this value (with default path length 100). For the treated mean, this maximum value is simply given by the largest unweighted treated imbalance,
\[
\lambda_1^{\max}
= \max_j
\left|
\frac{1}{n} \sum_{i=1}^{n} W_i X_{ij}
- \frac{1}{n} \sum_{i=1}^{n} X_{ij}
\right|.
\]

For lasso penalization, \balnet~also allows users to specify a target $\lambda_{\min}$ corresponding directly to a maximum allowable imbalance. The path solver then attempts to reach this tolerance and gracefully truncates the path if further reductions are infeasible\footnote{If the resulting imbalance remains unsatisfactory, users may consider augmenting the weighting estimator with an outcome model to account for the residual imbalance \citep{athey2018approximate}.}. By default, \balnet~standardizes the covariates, so $\lambda$ directly controls the maximum standardized mean difference.

Another benefit of path solvers is the efficient computation of cross-validation grids, as implemented in \cvbalnet. Direct approaches to cross-validation for regularized models fix a grid of penalty parameters and re-solve the optimization problem at each value (e.g., \citet{chattopadhyay2020balancing}). In contrast, the path-based approach computes the full sequence of regularization parameters at a computational cost that is often comparable to solving the optimization problem for a single value of $\lambda$ \citep{friedman2007pathwise}.

\section{Application: Pixel-level balancing} \label{sec:application}

We illustrate \balnet~using data from \citet{wu2023low}, who estimate the causal effect of past low-intensity fire on future high-intensity fires using NASA's MODIS instrument \citep{giglio2003enhanced}. Treatment status $W_i$ is equal to 1 for pixels that experienced a low-intensity fire in a given focal year and 0 otherwise, and the outcome is future high-intensity fire status. To construct a control group, the study conducts a synthetic control analysis \citep{abadie2010synthetic}, using covariate balancing to construct ATT weights such that pre-treatment fire history and geographic covariate trajectories match. 

We focus on a single focal year (2008) and land cover type (conifer forests), yielding a dataset with $n=141,780$ observations (4,466 treated) and $p=553$ covariates, measuring past fire history as well as topographic, meteorological, disturbance, and vegetation characteristics. The goal is to find weights that match the means of all 553 covariates in the synthetic control group to the corresponding treated-group means. The largest standardized mean difference in the unweighted data is approximately 470, occurring for snow water equivalent in September 2004.
We set the parameter \code{max.imbalance} to 0.05, targeting a maximum allowable standardized covariate imbalance of 0.05.
Internally, \balnet~adjusts the $\lambda$ sequence such that the resulting maximum standardized imbalance ranges from 470 down to the target 0.05 over \code{nlambda} $=100$ logarithmically spaced values.
\begin{CodeChunk}
\begin{CodeInput}
R> fit <- balnet(X, W, target = "ATT", max.imbalance = 0.05,
                 verbose = TRUE, num.threads = 4)
\end{CodeInput}
\end{CodeChunk}
The \code{verbose} option enables a progress bar that reports balance metrics along the $\lambda$ path. The \code{num.threads} argument controls the number of cores used by \adelie, and \balnet~for intermediate pre- and post-processing computations, including covariate standardization\footnote{Currently, only dense matrix input is supported, and standardization incurs a one-time $n \times p$ copy. The underlying \adelie~library supports zero-overhead lazy standardization via compile-time specialized matrix types (e.g., sparse and factor representations with optimized access operators and dot product routines \citep{yang2024fast}), enabling memory and computational savings. At the scales considered here, explicit standardization was sufficiently fast. Future work may exploit these specializations to scale to even larger datasets and richer feature representations.}.
Printing the returned object displays a summary of the path fit with default length \code{nlambda} $=100$. As the path is not truncated before reaching \code{nlambda} steps, further reductions in covariate imbalance are possible if desired.
\begin{CodeChunk}
\begin{CodeInput}
R> print(fit)
\end{CodeInput}
\begin{CodeOutput}
Call:  balnet(X = X, W = W, target = "ATT", max.imbalance = 0.05, verbose = TRUE, 
    num.threads = 4) 

Control (path: 100/100)
    Nonzero Avg|SMD|    Lambda
1         0  2.20052 467.85832
2         1  2.09474 426.58128
3         1  1.99770 388.94592
...
98       57  0.02622   0.06014
99       60  0.02415   0.05484
100      65  0.02222   0.05000
\end{CodeOutput}
\end{CodeChunk}
Motivated by standard diagnostics for covariate balance (e.g., \citet{austin2015moving, cohn2025ijmpr, greifer, keele2025balancing}), several commonly reported metrics are displayed. For a given $\lambda$, the estimated propensity scores targeting the control arm are $\hat e^{(0)}(X_i;\lambda)$ and the ATT weights are
$$
\hat \gamma_i(\lambda)
=
\frac{\hat e^{(0)}(X_i;\lambda)}{1 - \hat e^{(0)}(X_i;\lambda)}.
$$
Let $\bar X_{\text{target}}$ and $\bar \sigma_{\text{target}}$ denote the treated-group covariate means and standard deviations, and let
$$
\bar X^{(0)}(\lambda) = 
\frac{\sum_{i:W_i=0} \hat \gamma_i(\lambda) X_i}
{\sum_{i:W_i=0} \hat \gamma_i(\lambda)}
$$
denote the weighted control means. The $p$-vector of standardized mean differences (SMD) is
$$
\mathrm{SMD}(\lambda)
=
\frac{\bar X^{(0)}(\lambda)
-
\bar X_{\text{target}}
}{
\bar \sigma_{\text{target}}
}.
$$
Along with the number of nonzero coefficients, the print method also shows the average absolute SMD.
As a reminder, when using lasso penalization, \eqref{eq:cbps_kkt} implies that the maximum absolute standardized imbalance satisfies $|\mathrm{SMD}(\lambda)|_\infty \le \lambda$.

Calling the plot method displays two additional metrics along the $\lambda$ path.
\begin{CodeChunk}
\begin{CodeInput}
R> plot(fit)
\end{CodeInput}
\end{CodeChunk}
\begin{figure}[t]
    \centering
    \includegraphics[width=0.55\textwidth]{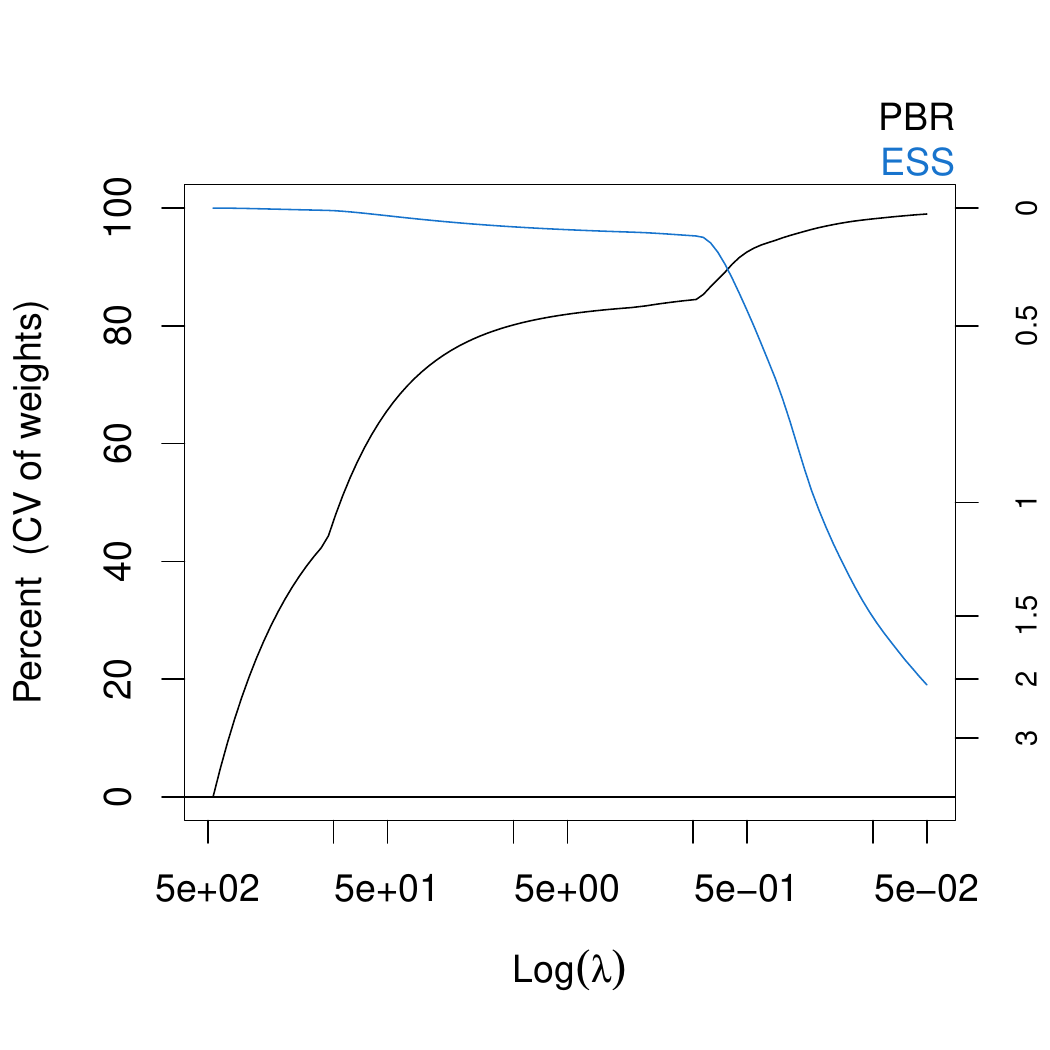}
    \caption{\footnotesize Regularization path for covariate balancing using data from \citet{wu2023low} for treatment year 2008 and conifer forests. The plot shows percentage bias reduction (PBR) and effective sample size (ESS) as functions of $\lambda$. The right-hand axis displays the coefficient of variation of the balancing weights.}
    \label{fig:path}
\end{figure} 
The percentage of bias reduction (PBR) is the reduction in average absolute SMD relative to the unweighted data (corresponding to $\lambda^{\max}$),
$$
\mathrm{PBR}(\lambda)
=
100 \times
\left(
1 -
\frac{
\mathrm{avg}|\mathrm{SMD}(\lambda)|
}{
\mathrm{avg}|\mathrm{SMD}(\lambda^{\max})|
}
\right).
$$
The effective sample size (ESS), expressed as a percentage, is
$$
\mathrm{ESS}(\lambda)
=
\frac{100}{n_0} \times
\frac{
\left(\sum_{i:W_i=0} \hat \gamma_i(\lambda)\right)^2
}{
\sum_{i:W_i=0} \hat \gamma_i(\lambda)^2
}, \quad n_0 = \sum_{i=1}^n (1-W_i).
$$
Figure \ref{fig:path} shows clear improvements in covariate balance, along with increased concentration of weights on a smaller subset of control units. The right axis in the figure displays the coefficient of variation of the weights $\hat \gamma_i(\lambda)$ via the identity $\sqrt{100 / \mathrm{ESS}(\lambda) - 1}$.

More granular information along the path can be obtained by supplying additional arguments to the plot method.
Supplying the \code{lambda} argument displays the individual SMDs using the estimated weights at a given $\lambda$. Below, we plot the SMDs for the 10 most imbalanced covariates at $\lambda_{\min}$ (which \code{lambda = 0} interpolates to, as in \glmnet).
\begin{CodeChunk}
\begin{CodeInput}
R> plot(fit, lambda = 0, max = 10)
\end{CodeInput}
\end{CodeChunk}
\begin{figure}[t]
    \centering
    \begin{subfigure}[b]{\textwidth}
        \centering
        \includegraphics[width=0.55\textwidth]{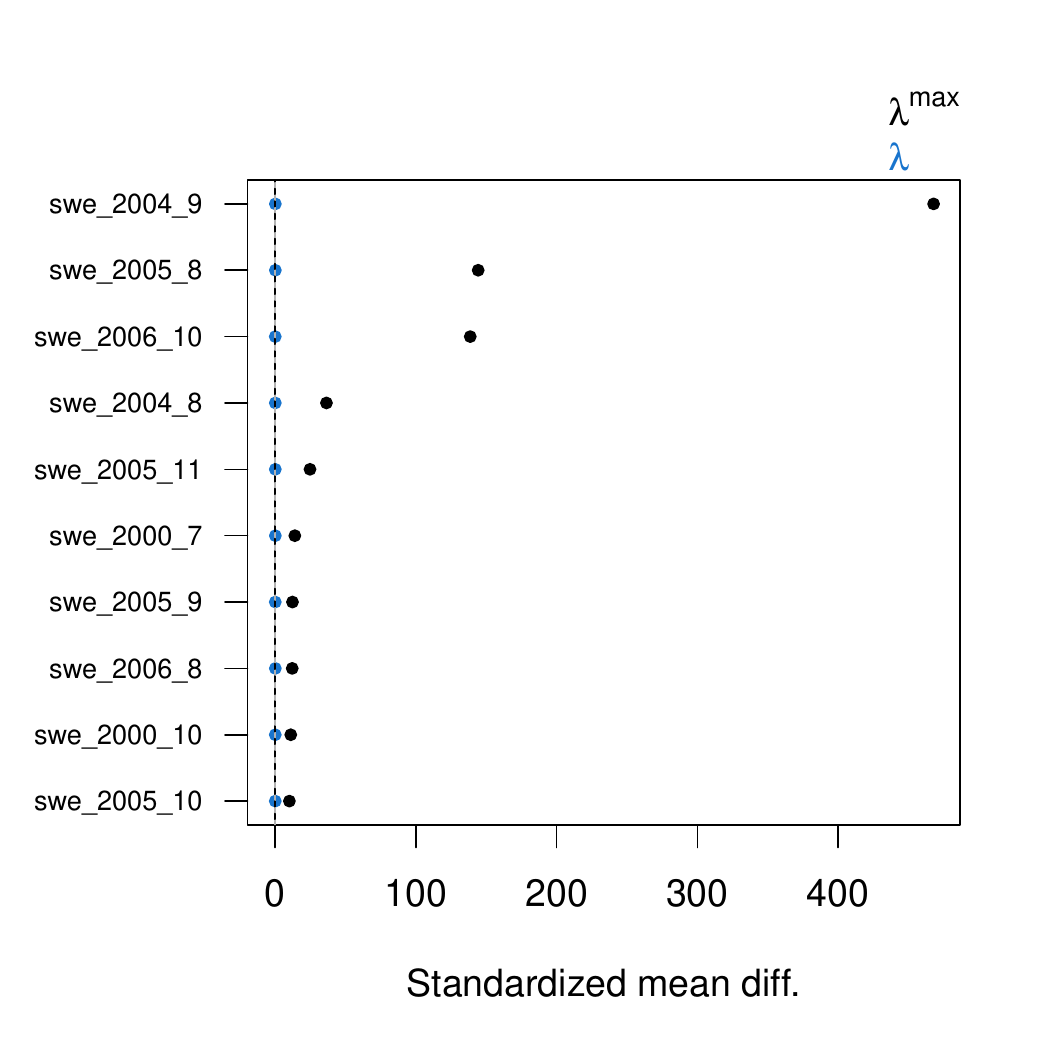}
        \caption[]
        {\footnotesize Standardized mean differences for the 10 most imbalanced covariates.}
        \label{fig:smd_top}
    \end{subfigure}
    \begin{subfigure}[b]{\textwidth}  
        \centering 
        \includegraphics[width=0.55\textwidth]{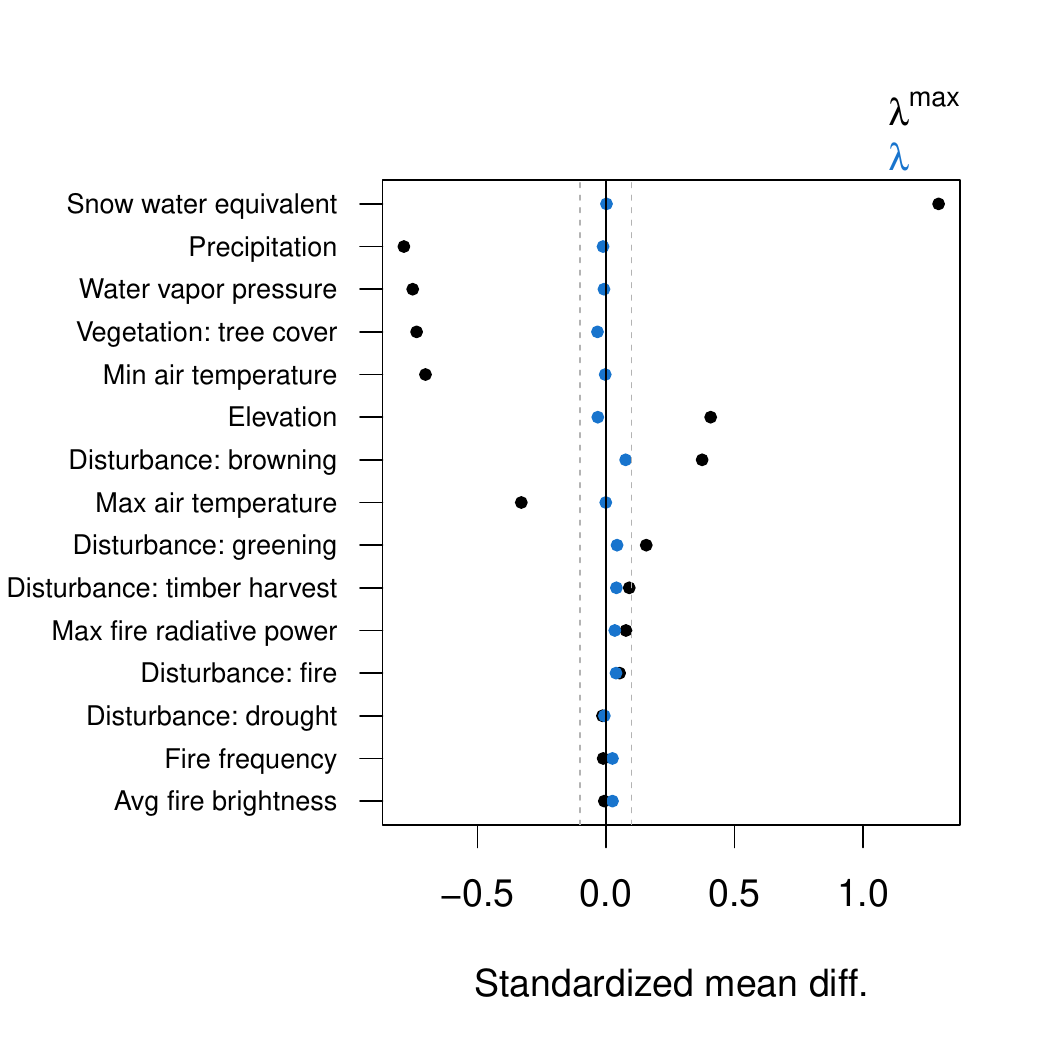}
        \caption[]
        {\footnotesize Standardized mean differences grouped by covariate category.}  
        \label{fig:smd_grp}
    \end{subfigure}
    \caption{\footnotesize Standardized mean differences before and after weighting, evaluated at $\lambda_{\min}$ from the path fit, with the unweighted data corresponding to $\lambda^{\max}$, using data from \citet{wu2023low} for treatment year 2008 and conifer forests.}
    \label{fig:smd}
\end{figure}
Figure \ref{fig:smd_top} shows the ten covariates with the largest unweighted imbalances, all related to snow water equivalent. Unweighted imbalances are shown as black dots, while weighted imbalances at the selected $\lambda$ are shown in color.
By construction, the absolute standardized mean difference for each covariate is bounded above by $\lambda$.
Visualizing all 500+ covariates is impractical, so we use the \code{groups} argument to \code{plot} to aggregate covariates by category. 
Figure \ref{fig:smd_grp} shows the resulting grouped imbalances.
\begin{CodeChunk}
\begin{CodeInput}
R> plot(fit, lambda = 0, groups = covariate.groups)
\end{CodeInput}
\end{CodeChunk}

\begin{figure}[t]
    \centering
    \includegraphics[width=0.55\textwidth]{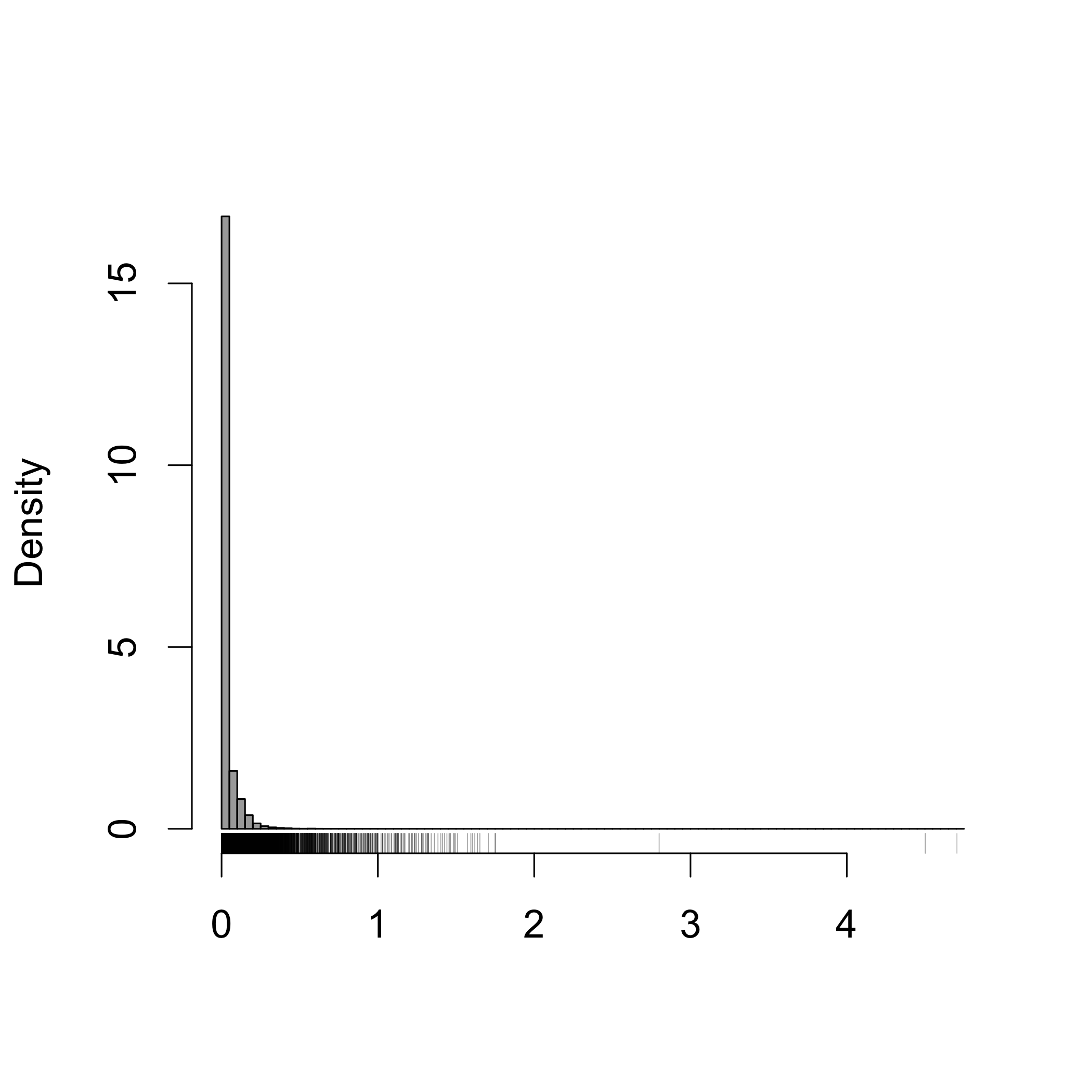}
    \caption{\footnotesize Estimated balancing ATT weights at $\lambda_{\min} = 0.05$.}
    \label{fig:weights}
\end{figure}

Finally, in addition to the standard \code{S3} methods for printing, plotting, and predicting, \balnet~provides an extractor for retrieving the balancing weights for any $\lambda$ on the path. We conclude by displaying the distribution of the estimated balancing weights at $\lambda_{\min}$ in Figure \ref{fig:weights}.

\FloatBarrier
\subsection{Timings}

To assess runtime, we conduct a small timing experiment using the data from this section\footnote{We do not include benchmarks against the \Rlang~packages mentioned in Section \ref{sec:intro} as they did not produce valid results under our experimental setup: \CBPS~did not complete within a one-hour time window, \RCAL~ returned results for only the first value of the regularization grid, with the remaining 99 entries being \code{NA}, and the remaining packages crashed on initialization.}.
We define a baseline dataset consisting of the first $n=100{,}000$ observations and $p=500$ covariates from the data described in the previous section. We then double both the sample size and the number of covariates by sampling with replacement from the baseline data. For each setting, we measure the time required for \balnet~to compute a regularization path targeting the ATT with maximum imbalance $\lambda_{\min} \in \{0.05, 0.01\}$. Table~\ref{tab:timing} reports runtimes obtained on an 8-core laptop with 24~GB of memory, using four cores for solver parallelism. The results suggest that runtime is more sensitive to the $\lambda$-sequence endpoints than to increases in the number of observations and covariates.

\begin{table}[ht]
\centering
\footnotesize
\begin{tabular}{lllccc}
\toprule
$n$ & $p$ & \code{max.imbalance} ($\lambda_{\min}$) &  Runtime (s) & $n,p$ scaling ($\times$) & $\lambda_{\min}$ scaling ($\times$) \\
\midrule
  100,000 & 500 & 0.05 & 4 & -- & --  \\
  200,000 & 1,000 & 0.05 & 13 & 3.2 & -- \\
  400,000 & 2,000 & 0.05 & 53 & 4.1 & -- \\
  \midrule
  100,000 & 500 & 0.01 & 43 & -- &  10.8 \\
  200,000 & 1,000 & 0.01 & 118 & 2.7 & 9.1 \\
  400,000 & 2,000 & 0.01 & 385 & 3.3 & 7.3 \\
\bottomrule
\end{tabular}
\caption{\footnotesize Runtimes (in seconds) and relative scaling when doubling both the number of observations and covariates, and tightening balance requirement. The number of cores used is 4.}
\label{tab:timing}
\end{table}

\section{Discussion}
Regularization is a practical necessity in many modern statistical settings and pathwise solutions are now a standard tool in regularized statistical learning \citep{friedman2007pathwise, glmnet}. We use pathwise regularization in a causal setting to target balancing weights for treatment effects under unconfoundedness via logistic covariate balancing loss functions. 

A natural direction for future work is extensions to additional causal estimands and settings.
In longitudinal settings with static treatment, such as panel data, balancing weights are often estimated along both time and unit dimensions, while dynamic treatment settings typically involve sequential optimization. 
An interesting question is whether pathwise GLM solvers can be adapted to accommodate these settings.

A direction for future algorithmic work is faster handling of regularization path truncation. In settings where exact balance is unattainable, there exists a value of $\lambda$ along the regularization path beyond which the penalized problem diverges.
Currently, \balnet~detects this point when \adelie~reaches the maximum number of coordinate descent iterations. At the default tolerance level and max iterations, this provides a robust indicator of the path endpoint in our experience, but it requires exhausting the full iteration budget.
A simple heuristic is to relax the tolerance (and thus solution precision) and halt when a step produces an explosive increase in the balance loss. Developing a more computationally efficient criteria for early detection of divergence remains for future work.

\section*{Acknowledgments}
We are grateful to Brandon de la Cuesta and Rina Park for helpful feedback.

\FloatBarrier
% \clearpage
\bibliographystyle{plainnat}
\bibliography{refs}
\end{document}